\documentstyle[rotating,epsf,preprint,tighten,prb,aps,floats]{revtex}
\begin{document}
\draft
\preprint{ }
\title{Photon side-bands in mesoscopics
\footnote{An article to commemorate Rolf Landauer's 70th birthday.}}
\author{A. P. Jauho}
\address{Mikroelektronik Centret, Technical University of Denmark, Bldg. 345e\\
DK-2800 Lyngby, Denmark}
\date{\today}
\maketitle
\begin{abstract}
This paper reviews several applications of photonic side-bands,
used by B{\"u}ttiker and Landauer in their theory of traversal
time in tunneling [Phys. Rev. Lett. {\bf 49}, 1739 (1982)], 
in transport and optics of mesoscopic systems.
Topics include generalizations of the transmission theory of
transport to time-dependent situations, optics and transport of mesoscopic
systems in THz electromagnetic fields, and phase-measurements
of photon-assisted tunneling through a quantum dot.
\end{abstract}
\medskip
\section{Introduction}
In 1982 Markus B{\"u}ttiker and Rolf Landauer published
a paper on traversal times in tunneling [\onlinecite{BL82}],
that rekindled the interest in an old topic and served as an inspiration
for much subsequent research.  Several articles
in this volume address the recent developments in this
field, which embraces a wide scope ranging from foundations
of quantum theory to practical questions concerning ultimate
speed limits of nanoelectronic components, a very characteristic
feature of Rolf Landauer's research.  
The point of the present article, however, is not
to participate in that particular discussion, but rather to use a technical
device
presented in the 1982-paper to analyze three other
physical systems all of which have connections to problems
Rolf Landauer has been active in.

The idea of B{\"u}ttiker and Landauer was to study
the sensitivity of tunneling transmission coefficient
through a potential barrier
to a time-periodic perturbation imposed on the barrier.
As the frequency of the modulation is varied, the inverse
of a cross-over frequency was identified as a characteristic
time-scale for tunneling. The solution to the time-dependent
Schr{\"o}dinger equation in the barrier region
was written, following Tien and Gordon [\onlinecite{TG63}], as
\begin{eqnarray}\label{psi(t)}
\psi_{\pm}(x,t,E) &=&
e^{\pm\kappa x}
e^{-iEt/\hbar}
\exp\left(-{iV_1\over\hbar\omega}\sin\omega t\right)\nonumber\\
&=&
e^{\pm\kappa x}
e^{-iEt/\hbar}
\left[\sum_{n=-\infty}^{n=+\infty}
J_n\left({V_1\over\hbar\omega}\right)
e^{-in\omega t}\right]\;,
\end{eqnarray}
where $V_1$ is the amplitude of the harmonic time-modulation,
$\omega$ is its frequency and 
$\kappa=\{2m[V_0-E]\}^{1/2}/\hbar$ with $V_0$
the barrier height.  The amplitudes of the side-bands
at energies $E\pm n\hbar\omega$
are given by the Bessel functions $J_n$.  Since the time-modulation
was introduced just as formal device to probe the energy sensitivity
of transmission, B{\"u}ttiker and Landauer were interested in 
the weak perturbation limit, $V_1/\hbar\omega\ll 1$.  In this paper,
on the other hand, 
we ask the question: What are the consequences of Eq.(\ref{psi(t)})
in physical situations where the external perturbation is {\it not}
weak?  Several examples will be discussed below.  In real physical
systems interactions (impurities, phonons, electron-electron collisions)
must be considered, and the interaction terms must be included
in the Hamiltonian.  It is often convenient to formulate the analysis
in the language of many-body formalism, and instead of working directly
with Eq.(\ref{psi(t)}), we prefer to use the spectral function,
which now reads
\begin{equation}\label{Gr}
A({\bf p},t,t') = \exp
\left[-i\int_{t'}^t dt_1\epsilon({\bf p},t_1)\right]\;.
\end{equation}
Note that the time-dependent perturbation requires one
to move away from the conventional energy representation,
$A({\bf p},E)=2\pi\delta(E-\epsilon({\bf p}))$.
Further, the above expression is a slight generalization of Eq.(1) in
that by an appropriate choice of the time-dependent
single-particle energy $\epsilon({\bf p},t)$ several
different physical systems can be addressed,
e.g., with $\epsilon({\bf p},t)=\epsilon({\bf p}-q{\bf A}(t))$,
where ${\bf A}(t)$ is the vector potential, a uniform time-dependent
electric field may be included, while setting
$\epsilon({\bf p},t)=p^2/2m+V_1\cos\omega t$ we return to 
situation discussed in Ref.[\onlinecite{BL82}], as
is readily verified by doing the integral in Eq.(\ref{Gr}),
and expanding the result in terms of Bessel functions,
in full analog with Eq.(\ref{psi(t)}).  We now proceed to
the applications.

\section{Time-dependent transport in mesoscopic systems}

In order that the consequences of Eqs.(\ref{psi(t)},\ref{Gr}) be
visible in an experiment at least the following two conditions
must be met.  First, the system must maintain a certain degree
of phase-coherence during a significant part of the transport
process.  Second, the time-variation must be different in
different parts of the system, and the particles must be
able to move between these regions.  This brings us in the
realm of mesoscopic systems.  Now, the transport in mesoscopic
systems can be analyzed extremely successfully in terms
of conductance formulas, pioneered by Landauer [\onlinecite{Lold}],
and their subsequent generalizations [\onlinecite{B86}].
Generically, one can express the conductance of a mesoscopic
system, coupled by ideal leads to external reservoirs, as
\begin{equation}\label{Landauer}
g = {e^2\over h} T(\epsilon_F) \;,
\end{equation}
where $T = |t|^2$ is the transmission coefficient, and $t$
is the complex transmission amplitude.   But
here we are addressing a problem going beyond the original
formulation: we have an external time-dependence acting on
the system.  Further, it would be desirable to include interactions within
the mesoscopic regions: these can be very important due to the
small number of charge carriers which implies less effective
screening.  

Recent years have witnessed a flurry of theoretical papers reporting
on generalizations of the original scattering-approach to 
interacting and/or time-dependent systems.  It would be beyond
the present purposes to provide a complete list of references,
instead we point to two chapters
[\onlinecite{APJ97},\onlinecite{B97}], representing complementary
views, in a forthcoming volume in Handbook of Semiconductors.
The formulation we adopt here focuses on the tunneling part of
the current\footnote{The current measured in the contacts also
contains contributions from displacement currents.  A low-frequency
theory for these, as well as long-range Coulomb forces is described
in Ref.[\onlinecite{B97}]}, and the current from the
left reservoir to the mesoscopic region
(we focus on a two-terminal geometry) can
be expressed as [\onlinecite{APJ97}] 
\begin{equation}\label{J(t)}
J_L(t) = - {2e\over\hbar}
\int_{-\infty}^t dt_1
\int {d\epsilon\over 2\pi}
{\rm ImTr}
\left\{ e^{-i\epsilon(t-t_1)}{\bf \Gamma}^L(\epsilon,t,t')
\left[{\bf G}^<(t,t_1) + f_L(\epsilon){\bf G}^r(t,t_1)\right]\right\}\;,
\end{equation}
Here the bold-face entities are matrices in the quantum numbers
specifying the states in the mesoscopic region; ${\bf\Gamma}^L$
is the coupling matrix to the left reservoir (the time-dependence
may be due to external gates which modify the potential barriers
between leads and the mesoscopic region) and the Green functions
${\bf G}^{<,r}$ must be calculated in the presence of the
coupling to the leads.  Typically, one would use the Dyson
equation to calculate ${\bf G}^r$ while ${\bf G}^<$ requires
the use of a quantum kinetic equation, e.g. the Keldysh equation.
Thus, Eq.(\ref{J(t)}) is  a {\it formal} expression for the 
time-dependent current, nevertheless it appears to form a
suitable starting point for further calculations, such as
those reported in 
Refs.[\onlinecite{S96a},\onlinecite{S96b},\onlinecite{Ng96}].

Many experiments focus on the average current and it is therefore
natural to ask whether Eq.(\ref{J(t)}) could be simplified in 
this case.  Indeed, one finds  
\begin{equation}\label{Jave}
\langle J_L(t) \rangle = - {2e \over \hbar}
\int {d\epsilon\over 2 \pi}
\left[f_L(\epsilon) - f_R(\epsilon)\right]
{\rm ImTr}
\left\{ 
{{\bf \Gamma}^L(\epsilon){\bf \Gamma}^R(\epsilon)\over
{\bf \Gamma}^L(\epsilon)+{\bf \Gamma}^R(\epsilon)}
\langle {\bf A}(\epsilon,t) \rangle
\right\}\;,
\end{equation}
where $f_{L/R}(\epsilon)=1/[1+\exp((\epsilon-\mu_{L/R})/kT)]$
are distribution functions describing the noninteracting
contacts with electro-chemical potential $\mu_{R/L}$, and
${\bf A}$ is an object closely related to the
retarded Green function [\onlinecite{JWM94}].
This expression is of the Landauer type: it expresses
the current as an integral over a weighted density of
states times the difference of the two contact occupation
factors.  
There is an important distinction, however:
the quantity in curly brackets does not involve, in general,
just the transmission coefficient but rather the Green
function for the fully interacting system, which
must be evaluated in the presence of interactions
({\it e.g.} electron-electron, electron-phonon,
and spin-flip).
The derivation of Eq.(\ref{Jave}) allows
arbitrary interactions in the
mesoscopic region, however the energy dependence of
the coupling matrices to right and left contacts must
be proportional to each other\footnote{This condition
is not too restrictive: in most cases the coupling matrices
are assumed to be constants (in lack of a detailed model)
and the conditions for the validity of Eq.(\ref{Jave})
are automatically satisfied.}.

Let us now apply these results to a simple example.
We consider a single, noninteracting state with
energy $\epsilon_0$ in the mesoscopic
region under the influence of a harmonically varying field
with amplitude $V_1$.
An explicit solution can readily be written down:

\begin{equation}\label{Aave}
\langle A(\epsilon,t)\rangle
= \sum_{k=-\infty}^{\infty}
J^2_k\left({V_1\over\hbar\omega}\right)
{\Gamma/2\over
\epsilon-\epsilon_0-k\omega+i\Gamma/2}\;.
\end{equation}
Combining Eqs.(\ref{Jave}) and the imaginary part of
(\ref{Aave}) we find that the
current can be written as
\begin{equation}\label{cur}
\langle J(t) \rangle =
{e\over h}\sum_{k=-\infty}^{\infty}\int d\epsilon
[f_L(\epsilon)-f_R(\epsilon)]
T(\epsilon-k\hbar\omega)J^2_k\left(
{V_1\over\hbar\omega}\right)\;,
\end{equation}
where $T(\epsilon)$ is the elastic transmission coefficient
through the mesocopic system.
The resulting low-field low-temperature conductance is then
\begin{equation}\label{Lac}
g_{\rm ac} = {e^2\over h}\sum_{k=-\infty}^\infty
T(\epsilon_F-k\hbar\omega)J^2_k\left({V_1\over\hbar\omega}\right)\;,
\end{equation}
which indeed appears as a natural generalization
of the standard conductance Eq.(\ref{Landauer}) to
the time-dependent situation.
These expressions bear a very close mathematical
resemblance to  the results
obtained by Tien and Gordon [\onlinecite{TG63}] and Tucker
[\onlinecite{Tuc79}], who found
\begin{equation}\label{Tucker}
\langle J(V_0,t) \rangle =
\sum_{k=-\infty}^\infty
J_k^2\left({V_1\over\hbar\omega}\right)I_{\rm dc}(V_0-k\hbar\omega/e)\;,
\end{equation}
{\it i.e.}, the rectified current of a system biased with
$V(t)=V_0 + V_1\cos\omega t$ is given as a sum of
dc-currents $I_{\rm dc}$ evaluated
at voltages shifted by integer multiples of photon energies.   
It is important
to note, however, 
that Eq.(\ref{cur}) was obtained as an explicit
calculation for a simple time-dependent resonant level keeping 
the coupling
to equilibrium contacts to all orders, while Refs.[\onlinecite{TG63},
\onlinecite{Tuc79}] only consider the lowest order coupling
between the different parts of the system, but do not
make other restrictive assumptions.
It would be interesting to learn more about the precise
interrelation of these approaches, in particular because
Eq.(\ref{Tucker}) has been very succesful in the analysis
of recent experiments on semiconductor superlattices
which are subjected to strong ac-fields
originating from  free electron lasers
[\onlinecite{Zeuner},\onlinecite{Gloria97,Wac97}].

\section{Linear optical absorption in THz-fields}

Our second application concerns the situation where the
system under investigation is placed in a strong THz-field (FIR-field)
and then its optical absorption coefficient is measured
with a weak probe field in the near infrared part
of the spectrum, corresponding to near band-edge
absorption in GaAs-based semiconductor systems.  As
we shall see, the photonic side-band structure
residing in Eq.(\ref{Gr}) leads to interesting, and
observable effects.  
The dominant features in the measured optical absorption
spectrum derive from two basic effects: those due to
a modified density of states, and those due to excitons, v.i.z.
interaction effects.
We analyze both of these effects; what makes our
discussion somewhat nonstandard is that we allow arbitrary
strength for the THz-field and thus go beyond
$\chi^{(3)}$-effects. 

Let us first consider the density of states effects.  In a
recent paper we have shown [\onlinecite{Kiddi}] that in the
noninteracting case the {\it time-dependent} absorption
coefficient $\alpha_T(\omega_l)$, where $\omega_l$ is
the frequency of the probe field, can be calculated from
\begin{equation}\label{abs}
\alpha_T(\omega_l)\simeq 
{2\pi^2\omega_l|d|^2\over c n \hbar}
\rho(T,\omega_l)\;,
\end{equation}
where $d$ is the dipole matrix element, $n$ is the refraction
coefficient, and the generalized density of states [\onlinecite{JJ96}] is
defined in terms of Eq.(\ref{Gr}),
\begin{equation}\label{rho}
\rho(T,\omega_l) = {1\over \pi}
\sum_{\bf k} A({\bf k},T,\omega_l)\;,
\end{equation}
where the Fourier-transform of Eq.(\ref{Gr}) is taken with respect
to $t-t'$, and $T=(t+t')/2$.  Already this simple result contains
some interesting physics.  In the dc-limit one recovers the standard
Franz-Keldysh absorption spectrum, {\it i.e.}, a finite (but exponentially
damped) absorption in the gap, and characteristic oscillations in
the band region.  In the time-dependent case, shown in Fig. 1,
a dynamical Franz-Keldysh
effect occurs [\onlinecite{JJ96},\onlinecite{Yac68}]: The overall
absorption edge experiences a blue-shift, and additional structure
appears, both in the gap region and in the continuum.
\begin{figure}
\begin{center}
\epsfxsize=10cm
\hspace{0.5cm}\epsfbox{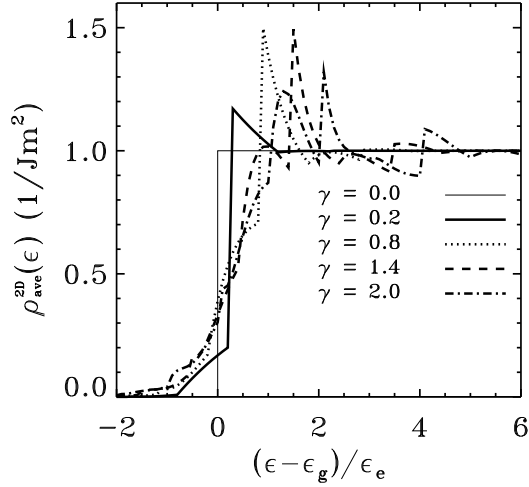}
\end{center}
\caption{ 
The time averaged GDOS for a 2D-system for a range of
THz field intensities, parametrized by 
$\gamma\equiv e^2E^2_{\rm THz}/4\hbar m^*\omega^3 = 
(0.2, 0.5, 0.8, 1.1, 1.4, 1.7, 2.0)$, as a function
of scaled energy ($\epsilon_e\equiv\hbar\omega$).
At low intensities one observes a Stark-like 
blue-shift of the band edge as well
as finite absorption within the band gap.
The blue-shift is given by $\epsilon_f\equiv \gamma\omega$,
and is physically interpreted as the average kinetic energy
of a classical charged particle with mass $m^*$ placed in an oscillating
electric field of frequency $\omega$ and strength $E_{\rm THz}$. 
With increasing 
intensity side bands emerge
at $\epsilon=\epsilon_g+\epsilon_f\pm 2\hbar\omega$.
(From Ref.[15])}
\label{2rf1}
\end{figure}

\begin{figure}[t]
\setlength{\unitlength}{1.0cm}
\begin{center}
\begin{picture}(9,11)
\epsfxsize=10cm
\put(-1,0.4){\epsfbox{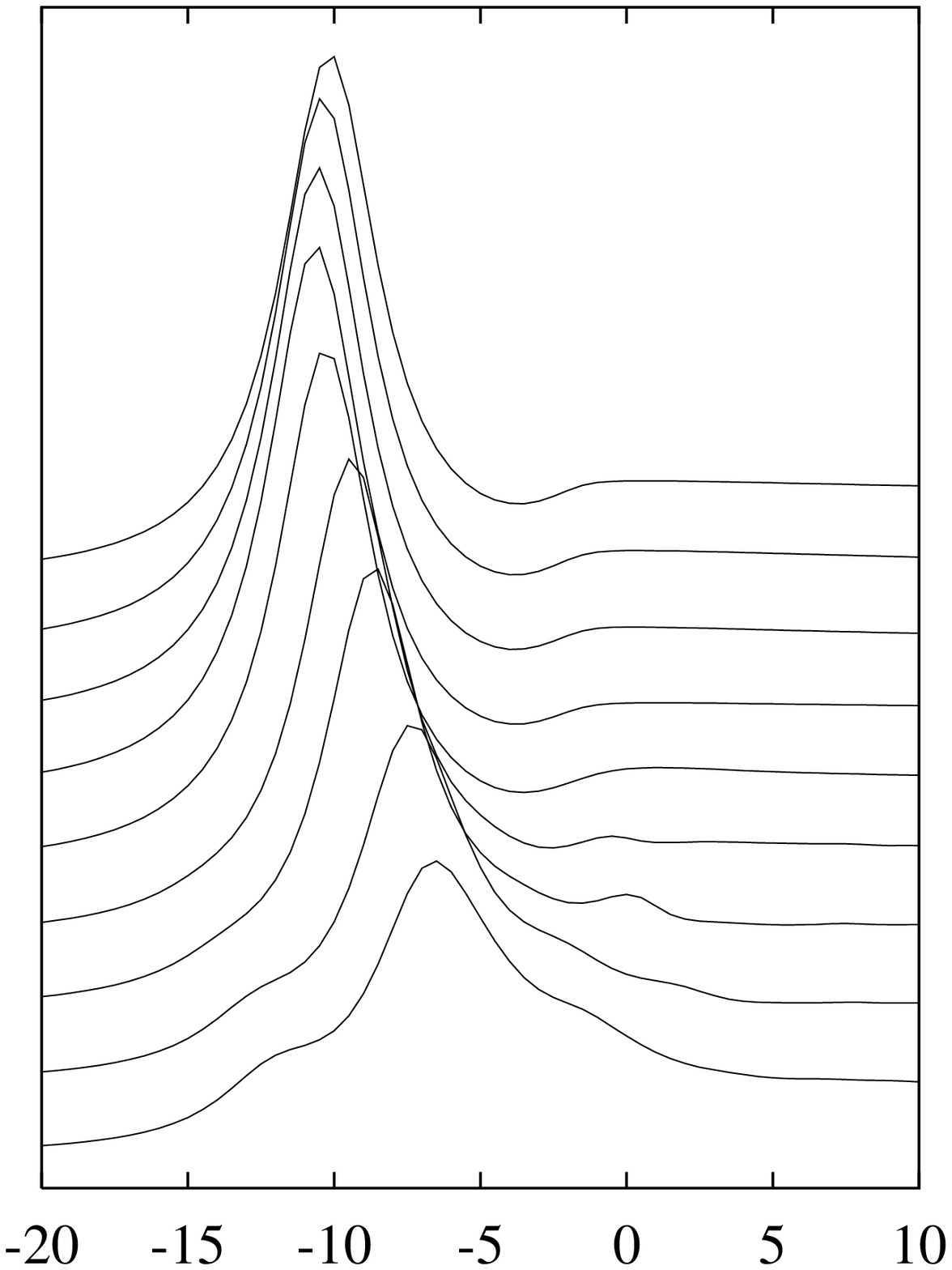}}
\put(3.5,0){$\hbar\omega_l-\epsilon_g$ (meV)}
\put(6.75,2.2){$\gamma = 6.0$}
\put(6.75,2.805){$\gamma = 4.5$}
\put(6.75,3.415){$\gamma = 2.5$}
\put(6.75,4.020){$\gamma = 1.5$}
\put(6.75,4.625){$\gamma = 0.5$}
\put(6.75,5.230){$\gamma = 0.1$}
\put(6.75,5.835){$\gamma = 0.06$}
\put(6.75,6.440){$\gamma = 0.03$}
\put(6.75,7.045){$\gamma = 0.0$}
\put(0.4,4.5){
\makebox(0.,0)[b]{\begin{sideways} Im$\chi_0(\omega_l)$ 
(arb. units)\end{sideways}}
}
\end{picture}
\end{center}
\caption{
Linear optical absorption in a quantum well for a range of
THz-intensities, parametrized by 
$\gamma=e^2E_{\rm THz}^2/4\hbar m^*\omega^3$.  The photon
energy is $\hbar\omega=2.5$ meV, below the ionization
threshold.  For small $\gamma$ the main feature is red-shifted,
while for increasing $\gamma$ a blue-shift, due to the
dynamical Franz-Keldysh effect, occurs.  Also, exciton
replicas at $\pm2\hbar\omega$ become visible.
(From Ref.[19])}
\end{figure}

Excitonic effects require a generalization of the analysis presented
above.  Rather than just focusing on the density of states,
one must evaluate  the susceptibility function,
$\chi^r(t,t')= -i\theta(t-t')\langle[P(t),P(t')]\rangle$, where
$P(t)$ is the polarization.  In the noninteracting limit one
can show that ${\rm Im}\chi^r_0(T,\omega_l)\propto\rho(T,\omega_l)$,
thus establishing a connection to Fig. 1.  The Coulomb interaction
between the electron and hole can be included by considering the
Bethe-Salpeter equation [\onlinecite{Haug}], which we have
generalized to include the THz-field nonperturbatively.  In the
photon side-band language the equation reads 
\begin{eqnarray}\label{BS}
\chi^r_n({\bf k},\omega_l+2n\omega)&=&
\chi^r_{0,n}({\bf k},\omega_l+2n\omega)\nonumber\\
&+&\sum_{n'}\chi^r_{0,n-n'}({\bf k},\omega_l+2(n+n')\omega)
\int{d{\bf k}\over(2\pi)^2} V(|{\bf k}-{\bf k'}|)
\chi^r_{n'}({\bf k},\omega_l+2n'\omega)\;,
\end{eqnarray}
where the Fourier representation is defined via
\begin{equation}
\chi^r(t,t')=\sum_n\int{d\omega\over 2\pi}
\chi^r_n(\omega)e^{i\omega_l(t-t')+in2\omega(t+t')}\;.
\end{equation}
This equation can be solved via standard numerical methods,
and the optical absorption coefficient is finally obtained
from ${\rm Im}\chi_0\equiv\sum_{\bf k}{\rm Im}\chi^r_{n=0}({\bf k},\omega_l)$.
Typical numerical results are shown in Fig. 2.
The theoretical predictions can be briefly summarized as
follows.  If the THz frequency is smaller than the frequency
corresponding to the ionization energy of the exciton
({\it i.e.} the energy difference corresponding to
$1s\to 2p$ transition), the ac Stark effect
leads to the red-shift seen at low intensity curves. However,
the dynamical Franz-Keldysh effect of Fig. 1 leads to a blue shift
which eventually overcomes the red-shift, and a net blue-shift
results.  Recent measurements performed at the UCSB Free Electron
Laser facility, to be fully described elsewhere [\onlinecite{Ken}],
are in very good agreement with these predictions.  On the
other hand, if the THz frequency is larger than the ionization
thresold, the ac Stark shift and dynamical Franz-Keldysh effect
work in unison, and a blue shift is always observed, both
theoretically and experimentally.

\section{Phase measurement of photon-assisted tunneling
through a quantum dot}

The basic conductance formula, Eq.(\ref{Landauer}), involves
the absolute square of the transmission amplitude.  The question
is then: Can one measure the {\it phase} of the transmission
amplitude? An affirmative answer was given by the recent groundbreaking
experiments of Yacoby {\it et al.} [\onlinecite{Yac95}] and
Schuster {\it et al.} [\onlinecite{Sch97}].
Their experimental protocol runs as follows:  A magneto-transport
measurement is performed on an Aharonov-Bohm ring with a quantum
dot fabricated in one of its arms.
If the quantum dot supports coherent transport, the transmission
amplitudes through the two arms interfere. 
A magnetic field induces
 a relative phase change, $2\pi \Phi/\Phi_0$, between the
two transmission amplitudes, $t_0$ and $t_{\rm QD}$,
leading to an oscillatory 
conductance $g(B) = (e^2/h){\cal T}(B)$, with 
\begin{equation}
{\cal T}(B)=
{\cal T}^{(0)} +
2{\rm Re}\{ t^*_0 
t_{\bf QD} e^{2\pi i \Phi/\Phi_0}\}
+...,
\label{Tring}
\end{equation}
where $\Phi$ is the flux threading 
the ring, $\Phi_0=hc/e$ is the flux quantum, 
and where the ellipsis represent higher harmonics due to multiple 
reflections. In the experiments, an oscillatory magnetoconductance of this form
was clearly observed thus demonstrating coherent transmission through
the dot [\onlinecite{Yac95,Sch97}]
Furthermore, controlling
the potential on the dot 
with a side-gate voltage,
allowed measurement of the 
phase shift of the transmission amplitude.  
The success of these experiments 
gave rise to a number of other works which 
concentrated on refining  the
interpretation of the experimental results
[\onlinecite{Levy,Yacoby2,Hackenbroich,Bruder}]. 
Yet, the experiments also suggest  application
to other phase-coherent transport 
processes. One particular example   which has been of considerable recent
interest, both experimentally  
[\onlinecite{Zeuner,Guim,Leo,Keay,Ooster}] 
and 
theoretically
[\onlinecite{Sokol,BS,Wagner,Gloria,Stoof,S96b}],
is photon-assisted tunneling. 
While photon-assisted tunneling (PAT) is intrinsically a 
coherent phenomenon, existing measurements
of PAT are insensitive to the phase of the transmitted electrons
and do not directly demonstrate coherence in the presence of
the time-dependent field.  
We have recently proposed [\onlinecite{JW}] a measurement of
photon-assisted tunneling through a quantum dot 
in the mesoscopic double-slit geometry described above. 
This is, in essence, a combination of the experiments of
Kouwenhoven {\it et al.} [\onlinecite{Leo,Ooster}] 
where a microwave modulated side-gate voltage gave rise
to photon-assisted tunneling through a quantum dot, and the
interference experiments 
of [\onlinecite{Yac95}] and [\onlinecite{Sch97}].

We focus on transport in the neighborhood of a single Coulomb
oscillation peak associated with a single nondegenerate electronic
level of the quantum dot [\onlinecite{Meirav}]. 
The effect of the ac side-gate voltage is
described entirely through the time-dependent energy of this level
\begin{equation}
\epsilon(t)=\epsilon_0(V_{\rm s})+V_1\cos \omega t\;, 
\label{energy}
\end{equation}
{\it i.e.}, precisely of the type considered in above.
Now we also emphasize that
the static energy of the level $\epsilon_0$ depends on the dc side-gate
voltage $V_{\rm s}$.
 All other levels on the
dot can be neglected provided  the ac amplitude, $V_1$,
 and the photon energy, $\hbar \omega$, are small
compared to the level spacing on the dot.

In the absence of an ac potential, a suitable model for 
the transmission amplitude 
$t_{\rm QD}(\epsilon)$ 
through the dot is the Breit-Wigner form,
\begin{equation}
t_{\rm QD}(\epsilon)={{-i\sqrt{\Gamma_L \Gamma_R}} \over 
{\epsilon - \epsilon_0(V_{\rm s}) + i\Gamma/2}}\;,
\label{tqdstatic}
\end{equation}
where 
$\Gamma=\Gamma_L+\Gamma_R$
is the full width at half maximum of the resonance 
on the dot due to tunneling to the left and right leads.
Eq. (\ref{tqdstatic}) implies 
a continuous phase accumulation  of $\pi$ 
in the transmission amplitude as the Coulomb
blockade peak is traversed. (Note that 
the Breit-Wigner form is exact for a noninteracting system with $\Gamma$
independent of energy.)

In the dynamic case, the simple Breit-Wigner description
must be generalized, and the object to evaluate
is the $S$-Matrix element [\onlinecite{JWM94,WJW}]. 
Provided interactions in the leads can be neglected, the 
elastic transmission amplitude $t_{\rm QD}(\epsilon)$
can be written as the energy conserving part
of the $S$-Matrix between the left lead and the right lead
\begin{equation}
\lim_{\epsilon' \rightarrow \epsilon} 
{\langle \epsilon', R | {\cal S} | \epsilon, L \rangle} =
\delta(\epsilon' - \epsilon) t_{\rm QD}(\epsilon)\;.
\label{Svst}
\end{equation}
\begin{figure}
\begin{center}
\epsfxsize=8.0cm
\hspace{2cm}
\epsfbox{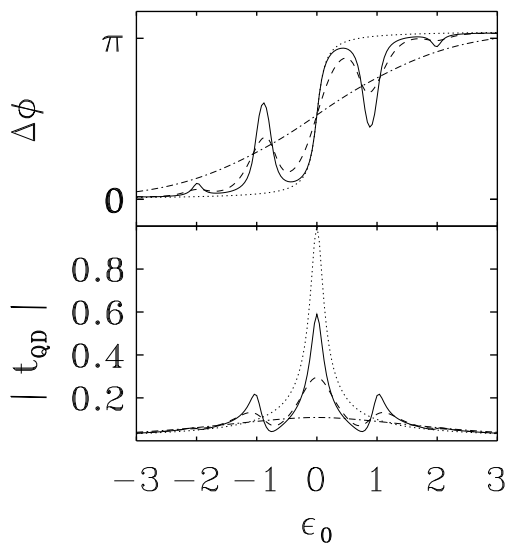}
\vspace{1.5cm}
\end{center}
\caption{Temperature dependence of the phase shift $\Delta\phi$ (top panel) and
the square of the amplitude (bottom) of 
${t}_{\rm QD}$. 
The level-width is $\Gamma/2=0.1$, in terms of which the other
parameters are $V_1=1.0$, $\omega = 1.0$, and 
$T = 0$ (solid line), $0.1$ (dashed line),  
$0.5$ (dash-dotted line).  For comparison, the $T=0$
time-independent results are shown as dots.
(From Ref.[39])  } 
\label{f3}
\end{figure}

The $S$-Matrix is simply related to the retarded Green function
of the level on the dot, including both tunneling to the leads and the ac
potential [\onlinecite{WJW}], and we find [\onlinecite{JW}]
\begin{equation}
t_{\rm QD}(\epsilon) = -i \sqrt{\Gamma_L \Gamma_R} 
\langle A(\epsilon,t)\rangle\;.
\label{tvsA}
\end{equation}
In the spirit of the Breit-Wigner transmission amplitude
we can use the noninteracting
$\langle A(\epsilon,t)\rangle$ given by Eq.(\ref{Aave})
in further calculations.
At finite temperatures one must compute
${t}_{\rm QD}=\int\! 
d\epsilon(-\partial f_0/\partial \epsilon)
{t}_{\rm QD}(\epsilon)$ where $f_0(\epsilon)$ is the Fermi function, 
and the final result is
\begin{equation}
{t}_{\rm QD} =\left(- {\Gamma\over 4\pi T}\right)
\sum_{k=-\infty}^{\infty} J_k^2(V_{\rm ac}/\hbar\omega)
\psi'[{1\over 2}-{i\over 2\pi T}(\mu-\epsilon_0(V_{\rm s})-
k\hbar\omega+i{\Gamma\over 2})]\;,
\label{mainresult}
\end{equation}
where $\psi'$ is the derivative of the digamma function, and
$\mu$ is the chemical potential in the leads.

We emphasize
that a conventional conductance measurement would yield information
only about the time average of
{\it the square} of the
transmission amplitude, and the double-slit geometry
of Ref.[\onlinecite{Yac95,Sch97}]  
is necessary in order to probe the phase.
Figure \ref{f3} shows the computed magnitude of 
${t}_{\rm QD}$ (bottom)
and its phase (top), as a function of the level energy
$\epsilon_0(V_{\rm s})$.
As compared to the time-independent case (shown as a dotted line),
several features are noteworthy.  The magnitude of
${t}_{\rm QD}$ 
shows photonic side-bands, reminiscent of those seen in transmission
through a microwave modulated quantum dot [\onlinecite{Leo}].  
However, there is an 
important difference from the usual case of photon-assisted
tunneling. The amplitude of the Aharonov-Bohm oscillation is 
sensitive only to the time average of 
the transmission amplitude
$t_{\rm QD}$.
Hence only elastic transmission through the dot contributes,
{\it i. e.}, the net number of photons absorbed from the ac
field must be zero. The sideband at say $\epsilon = \epsilon_0(V_s) -
\hbar \omega$ corresponds to a process in which an electron
first absorbs a photon to become resonant at energy $\epsilon_0(V_s)$,
and subsequently reemits the photon to return to its original energy.
In Ref.[\onlinecite{JW}] we have studied the phase as a function
of the strength of the time-dependent modulation, and find that
it is possible to quench the main transmission peak, or change
the sign of the slope or the phase at resonance by adjusting the
ratio $V_1/\hbar\omega$ to coincide with a zero of the Bessel
function $J_0$.  

\begin{acknowledgements}
The work described in this paper was carried in collaboration with
Ned Wingreen (Sections 2 and 4), Yigal Meir (Section 2), 
and Kristinn Johnsen (Section 3), to whom the author
expresses his gratitude. A guest professorship from the Iberdrola
Foundation provided delightful working conditions at ICMM in 
Madrid during the
completition of this paper.  
The author is grateful to Gloria Platero and Ram{\'o}n Aguado
for useful comments on the manuscript.
He also acknowledges the receipt of
several notes, letters, corrections to incorrectly (un)quoted
papers, and emails from Rolf Landauer during the
last ten years or so, most of which, while perhaps critical in nature,
at the same time were
constructive and aimed at clarifying positions in difficult issues.
Perhaps it is appropriate to end in quoting an excerpt from
an email received in May 1988:
``.. I am a veteran of many controversial views and fields;
the truth in science is not reached easily, otherwise we would
be unemployed. Disagreement, in print, does not mean lack of
respect, and can be constructive ..... 
I doubt if we will come into complete intellectual equilibrium,
and do not believe that is essential.''
\end{acknowledgements}

\end{document}